\pdfoutput=1
\documentclass[12pt,letter]{article}
\usepackage{ifthen} 
\usepackage{booktabs}
\usepackage{multicol}
\usepackage{multirow}

\usepackage[style = ieee]{biblatex}
\newboolean{pdflatex}
\setboolean{pdflatex}{true} 
\newboolean{articletitles}
\setboolean{articletitles}{true} % False removes titles in references
\newboolean{uprightparticles}
\setboolean{uprightparticles}{false} %True for upright particle symbols
\def\paperauthors{HEV Collaboration} % Leave as is for PAPER, CONF and FIGURE
\def\paperasciititle{The HEV Ventilator Proposal} % Set ASCII title here !! MAKE sure it's only ASCII characters !!
\def\papertitle{The HEV Ventilator Proposal } % Latex formatted title
\def\paperkeywords{{Ventilator},{COVID-19}} % 
\def\papercopyright{\the\year\ CERN } % new since 9/Apr/2018
\def\paperlicence{CC-BY-4.0 licence}
\def\paperlicenceurl{https://creativecommons.org/licenses/by/4.0/}
\usepackage{url}

\usepackage[top=1in, bottom=1.25in, left=1in, right=1in]{geometry}

\usepackage{microtype}
\usepackage{lineno}  % for line numbering during review
\usepackage{xspace} % To avoid problems with missing or double spaces after
\usepackage{caption} %these three command get the figure and table captions automatically small

\usepackage{subfigure}

\usepackage{graphicx}  % to include figures (can also use other packages)
\usepackage{color}
\usepackage{colortbl}
\graphicspath{{./figs/}} % Make Latex search fig subdir for figures
\usepackage{amsmath} % Adds a large collection of math symbols
\usepackage{amssymb}
\usepackage{amsfonts}
\usepackage{upgreek} % Adds in support for greek letters in roman typeset

\newcommand*\patchAmsMathEnvironmentForLineno[1]{%
\expandafter\let\csname old#1\expandafter\endcsname\csname #1\endcsname
\expandafter\let\csname oldend#1\expandafter\endcsname\csname
end#1\endcsname
 \renewenvironment{#1}%
   {\linenomath\csname old#1\endcsname}%
   {\csname oldend#1\endcsname\endlinenomath}%
}
\newcommand*\patchBothAmsMathEnvironmentsForLineno[1]{%
  \patchAmsMathEnvironmentForLineno{#1}%
  \patchAmsMathEnvironmentForLineno{#1*}%
}
\AtBeginDocument{%
\patchBothAmsMathEnvironmentsForLineno{equation}%
\patchBothAmsMathEnvironmentsForLineno{align}%
\patchBothAmsMathEnvironmentsForLineno{flalign}%
\patchBothAmsMathEnvironmentsForLineno{alignat}%
\patchBothAmsMathEnvironmentsForLineno{gather}%
\patchBothAmsMathEnvironmentsForLineno{multline}%
\patchBothAmsMathEnvironmentsForLineno{eqnarray}%
}

\usepackage{hyperxmp}

\usepackage[pdftex,
            pdfauthor={\paperauthors},
            pdftitle={\paperasciititle},
            pdfkeywords={\paperkeywords},
            pdfcopyright={Copyright (C) \papercopyright},
            pdflicenseurl={\paperlicenceurl}]{hyperref}
\usepackage[colorinlistoftodos,textsize=scriptsize]{todonotes}

\usepackage[bottom,flushmargin,hang,multiple]{footmisc}

\usepackage[all]{hypcap} % Internal hyperlinks to floats.

\usepackage{xspace} 
\usepackage{upgreek}

\def\MagUp {\mbox{\em Mag\kern -0.05em Up}\xspace}

\ifthenelse{\boolean{uprightparticles}}%
{

 \def\PDelta      {\ensuremath{\Delta}\xspace}                 
 \def\PXi         {\ensuremath{\Xi}\xspace}                 
 \def\PLambda     {\ensuremath{\Lambda}\xspace}                 
 \def\PSigma      {\ensuremath{\Sigma}\xspace}                 
 \def\POmega      {\ensuremath{\Omega}\xspace}                 
 \def\PUpsilon    {\ensuremath{\Upsilon}\xspace}

 \def\PB      {\ensuremath{\mathrm{B}}\xspace}                 
                  
 \def\PD      {\ensuremath{\mathrm{D}}\xspace}

 \def\PK      {\ensuremath{\mathrm{K}}\xspace}

 \def\Pi      {\ensuremath{\mathrm{i}}\xspace}

 \def\Ps      {\ensuremath{\mathrm{s}}\xspace}

 \def\thebaroffset{0.0em}
}
{

 \mathchardef\PDelta="7101
 \mathchardef\PXi="7104
 \mathchardef\PLambda="7103
 \mathchardef\PSigma="7106
 \mathchardef\POmega="710A
 \mathchardef\PUpsilon="7107
                  
 \def\PB      {\ensuremath{B}\xspace}                 
                  
 \def\PD      {\ensuremath{D}\xspace}

 \def\PK      {\ensuremath{K}\xspace}

 \def\Pi      {\ensuremath{i}\xspace}

 \def\Ps      {\ensuremath{s}\xspace}

 \def\thebaroffset{0.18em}
}
\newcommand{\offsetoverline}[2][\thebaroffset]{\kern #1\overline{\kern -#1 #2}}%

\makeatletter
\ifcase \@ptsize \relax% 10pt
  \newcommand{\miniscule}{\@setfontsize\miniscule{4}{5}}% \tiny: 5/6
\or% 11pt
  \newcommand{\miniscule}{\@setfontsize\miniscule{5}{6}}% \tiny: 6/7
\or% 12pt
  \newcommand{\miniscule}{\@setfontsize\miniscule{5}{6}}% \tiny: 6/7
\fi
\makeatother

\DeclareRobustCommand{\optbar}[1]{\shortstack{{\miniscule (\rule[.5ex]{1.25em}{.18mm})}
  \\ [-.7ex] $#1$}}

\def\squark    {{\ensuremath{\Ps}}\xspace}

\def\KorKbar {\kern \thebaroffset\optbar{\kern -\thebaroffset \PK}{}\xspace}

\def\D       {{\ensuremath{\PD}}\xspace}

\def\DorDbar {\kern \thebaroffset\optbar{\kern -\thebaroffset \PD}\xspace}

\def\Dp      {{\ensuremath{\D^+}}\xspace}
\def\Dm      {{\ensuremath{\D^-}}\xspace}

\def\DpDm    {\ensuremath{\Dp {\kern -0.16em \Dm}}\xspace}

\def\B       {{\ensuremath{\PB}}\xspace}

\def\BorBbar {\kern \thebaroffset\optbar{\kern -\thebaroffset \PB}\xspace}

\def\Bd      {{\ensuremath{\B^0}}\xspace}

\def\BdorBdbar {\kern \thebaroffset\optbar{\kern -\thebaroffset \Bd}\xspace}

\def\Bs      {{\ensuremath{\B^0_\squark}}\xspace}

\def\BsorBsbar {\kern \thebaroffset\optbar{\kern -\thebaroffset \Bs}\xspace}

\def\Y#1S{\ensuremath{\PUpsilon{(#1S)}}\xspace}

\def\LorLbar     {\kern \thebaroffset\optbar{\kern -\thebaroffset \PLambda}\xspace}

\def\AT#1     {\ensuremath{A_{\mathrm{T}}^{#1}}\xspace}           % 2

\def\C#1      {\ensuremath{\mathcal{C}_{#1}}\xspace}                       % 9
\def\Cp#1     {\ensuremath{\mathcal{C}_{#1}^{'}}\xspace}                    % 7
\def\Ceff#1   {\ensuremath{\mathcal{C}_{#1}^{\mathrm{(eff)}}}\xspace}        % 9  
\def\Cpeff#1  {\ensuremath{\mathcal{C}_{#1}^{'\mathrm{(eff)}}}\xspace}       % 7
\def\Ope#1    {\ensuremath{\mathcal{O}_{#1}}\xspace}                       % 2
\def\Opep#1   {\ensuremath{\mathcal{O}_{#1}^{'}}\xspace}                    % 7

\newcommand{\aunit}[1]{\ensuremath{\text{\,#1}}}       
\newcommand{\tev}{\aunit{Te\kern -0.1em V}\xspace}
\newcommand{\gev}{\aunit{Ge\kern -0.1em V}\xspace}
\newcommand{\mev}{\aunit{Me\kern -0.1em V}\xspace}
\newcommand{\kev}{\aunit{ke\kern -0.1em V}\xspace}
\newcommand{\ev}{\aunit{e\kern -0.1em V}\xspace}
 
\newcommand{\mevc}{\ensuremath{\aunit{Me\kern -0.1em V\!/}c}\xspace}
\newcommand{\gevc}{\ensuremath{\aunit{Ge\kern -0.1em V\!/}c}\xspace}
\newcommand{\mevcc}{\ensuremath{\aunit{Me\kern -0.1em V\!/}c^2}\xspace}
\newcommand{\gevcc}{\ensuremath{\aunit{Ge\kern -0.1em V\!/}c^2}\xspace}
\def\gsim{{~\raise.15em\hbox{$>$}\kern-.85em
          \lower.35em\hbox{$\sim$}~}\xspace}
\def\lsim{{~\raise.15em\hbox{$<$}\kern-.85em
          \lower.35em\hbox{$\sim$}~}\xspace}

\def\tell1  {TELL1\xspace}
\def\ukl1   {UKL1\xspace}

\usepackage{cleveref}

\usepackage{longtable} % only for template; not usually to be used in PAPERs
\usepackage{siunitx}

\begin{document}
\renewcommand{\thefootnote}{\fnsymbol{footnote}}
\setcounter{footnote}{1}
%\onecolumn
% $Id: title-LHCb-PAPER.tex 122889 2018-08-17 17:59:55Z pkoppenb $
% ===============================================================================
% Purpose: LHCb-PAPER journal paper title page template
% Author: 
% Created on: 2010-09-25
% ===============================================================================

%%%%%%%%%%%%%%%%%%%%%%%%%
%%%%%  TITLE PAGE  %%%%%%
%%%%%%%%%%%%%%%%%%%%%%%%%
\begin{titlepage}
\pagenumbering{roman}

% Header ---------------------------------------------------
\vspace*{-1.5cm}
\centerline{\large EUROPEAN ORGANIZATION FOR NUCLEAR RESEARCH (CERN)}
\vspace*{1.5cm}
\noindent
\begin{minipage}{0.6\textwidth}
\includegraphics[width=.25\textwidth]{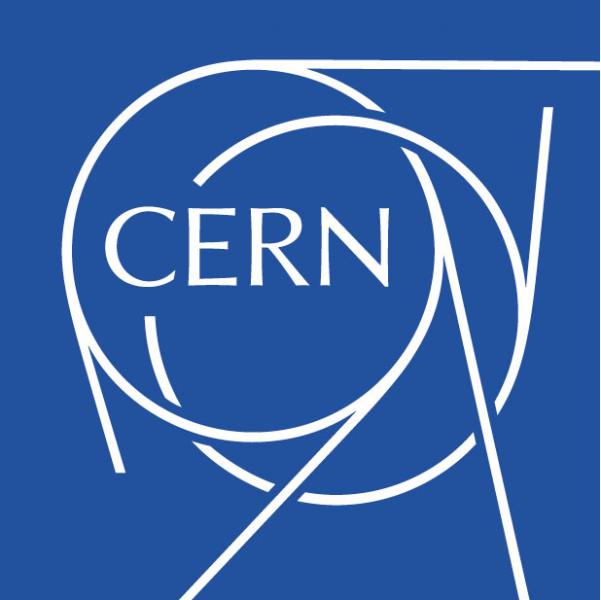}
\includegraphics[width=.4\textwidth]{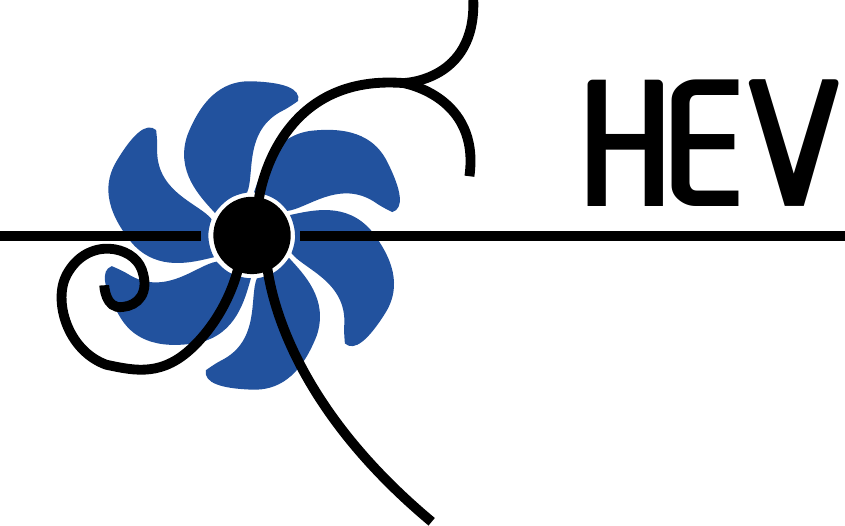}
\end{minipage}
\begin{minipage}{0.4\textwidth}
\begin{flushright}
CERN-EP-TECH-NOTE-2020-01\\
\today
\end{flushright}
\end{minipage}\\
%\begin{tabular*}{\linewidth}{lc@{\extracolsep{\fill}}r}
%\ifthenelse{\boolean{pdflatex}}% Logo format choice
%{\vspace*{-1.5cm}\mbox{\!\!\!\includegraphics[width=.34\textwidth]{figs/cern-hev-logo}} & & CERN-EP-TECH-NOTE-001}\\
%{ & & \today} % Date - Can also hardwire e.g.: 23 March 
%\end{tabular*}

\vspace*{4.0cm}
{\normalfont\bfseries\boldmath\huge
\begin{center}
% DO NOT EDIT HERE. Instead edit macro in main.tex to keep metadata correct
  \papertitle 
\end{center}
}
\vspace*{2.0cm}
% Authors -------------------------------------------------
\begin{center}
%In the footnote, replace 'paper' by 'Letter' in case of submission to PRL or PLB 
% Edit macro in main.tex to keep metadata correct
\paperauthors{\footnote{Authors are listed at the end of this paper.}}
\end{center}
\vspace{\fill}
% Abstract -----------------------------------------------
\begin{abstract}
  \noindent
\end{abstract}
\vspace*{2.0cm}
We propose the design of a ventilator which can be easily manufactured and integrated into the hospital environment to support COVID-19 patients. The unit is designed to support standard ventilator modes of operation, most importantly PRVC (Pressure Regulated Volume Control) and SIMV-PC (Synchronised Intermittent Mandatory Ventilation) modes. The unit is not yet an approved medical device and is in the concept and prototyping stage. It is presented here to invite fast feedback for development and deployment in the face of the COVID-19 pandemic.
\begin{center}

\end{center}
\vspace{\fill}
{\footnotesize 
% Edit macro in main.tex to keep metadata correct
\centerline{\copyright~\papercopyright. \href{\paperlicenceurl}{\paperlicence}.}}
\vspace*{2mm}
\end{titlepage}
\newpage
\setcounter{page}{2}
\mbox{~}
%\twocolumn
\renewcommand{\thefootnote}{\arabic{footnote}}
\setcounter{footnote}{0}
\tableofcontents
\cleardoublepage
\pagestyle{plain} % restore page numbers for the main text
\setcounter{page}{1}
\pagenumbering{arabic}
%\linenumbers

\section{Motivation}
\label{sec:motivation}

The worldwide medical community currently faces a critical shortage of medical equipment to address the COVID-19 pandemic~\cite{axios,nytimes,musk}.  In particular this is the case for ventilators, which are needed during COVID-19 related treatment at onset, during the intensive care phase and during the very extended recovery times.  Companies are scaling up production~\cite{scaling}, but this will not be sufficient to meet the demand according to the current forecasts. There is a wide spectrum of devices, ranging from highly sophisticated through to simpler units~\cite{oxylator,emergency_ventilator} useful in the milder phases of illness.  There is already a large number of proposals circulating for devices which can be quickly manufactured cheaply and on large scale~\cite{mvm,mvmurl,rice,ventilad,oxvent,belgian}.

We propose here a ventilator design to be integrated into a hospital environment.  The HEV (High Energy physics community Ventilator) concept is based on components which are simple and cheap to source, complies with hospital standards for external connections and operating modes, and supports the most requested operation modes.  The starting point of the design is the set of MHRA (Medicines and Healthcare products Regulatory Agency) guidelines provided by the UK government regarding Rapidly Manufactured Ventilator Systems~\cite{mhra}.  The proposal here is, at this stage, not a medically approved system and will need a process of verification with medical experts.  However, in the interests of rapid development the concept is presented to generate feedback, corrections, and support as the project progresses.  A demonstrator has been built and the design is now in the prototyping stage.

\section{Modes of Operation; overview}

Patient management during COVID-19 faces serious issues of lung damage, and the ventilators must be able to handle situations of rapidly changing lung compliance, and potential collapse and consolidation. The driving pressure of the ventilator is a crucial factor for patient outcomes~\cite{hamilton}. In particular, when a low tidal volume is used, the driving pressure is an important variable to monitor to assess the risk of hospital mortality.

In light of the extreme importance of the pressure monitoring, the HEV ventilator will target pressure controlled modes. This will include: PRVC (Pressure Regulated Volume Control) mode, SIMV-PC (Synchronised Intermittent Mandatory Ventilation); and in addition a basic mode of operation: CPAP (Continuous Positive Airway Pressure). The HEV design also provides PEEP (Positive End-Expiratory Pressure), which is not a ventilatory mode in itself but is designed to support steady low positive pressure to the lungs. The PRVC mode, which is standard for commercial ventilators, aims to set the tidal volume at the lowest possible airway pressure.  In the case where the tidal volume is not achieved at a particular pressure setting, due to changes in the patient's airway resistance or lung compliance; this can then be gradually adjusted. SIMV-PC mode will allow the patient to take spontaneous breaths, and will assist the breathing when the spontaneous breath is taken.  This mode uses an additional sensor for the detection of the negative pressure initiated by the patient breath.  If the patient respiratory rate does not achieve the target value, additional mechanical ventilation is provided by the unit.

The HEV ventilator will also be capable of a basic non-invasive operation mode where a fixed pressure is made available to the patient.  Although international definitions vary, this corresponds to the CPAP definition from the MHRA documentation~\cite{mhra}.  In all modes of operation, PEEP will be available, which is important for patient management to avoid alveolar collapse.

Note that the ventilator design outlined here is not intended to replace the high-end devices needed for the most intense phase of treatment, but should be appropriate and useful in the hospital environment for milder symptoms or long term care and recovery.

\section{Conceptual Design}

We describe here the conceptual layout of the system. The targeted modes of operation, as explained above, are principally PRVC, SIMV-PC and CPAP, as defined in the MHRA documentation~\cite{mhra}. The design has the patient safety built-in as a priority, so that all failure modes revert to a situation which prioritises patient safety. In particular, if the patient stops breathing in pressure support mode, the ventilator fail-safes automatically onto mandatory ventilation.

\begin{figure}[ht]
\begin{center}
  
  \includegraphics[width=0.95\linewidth]{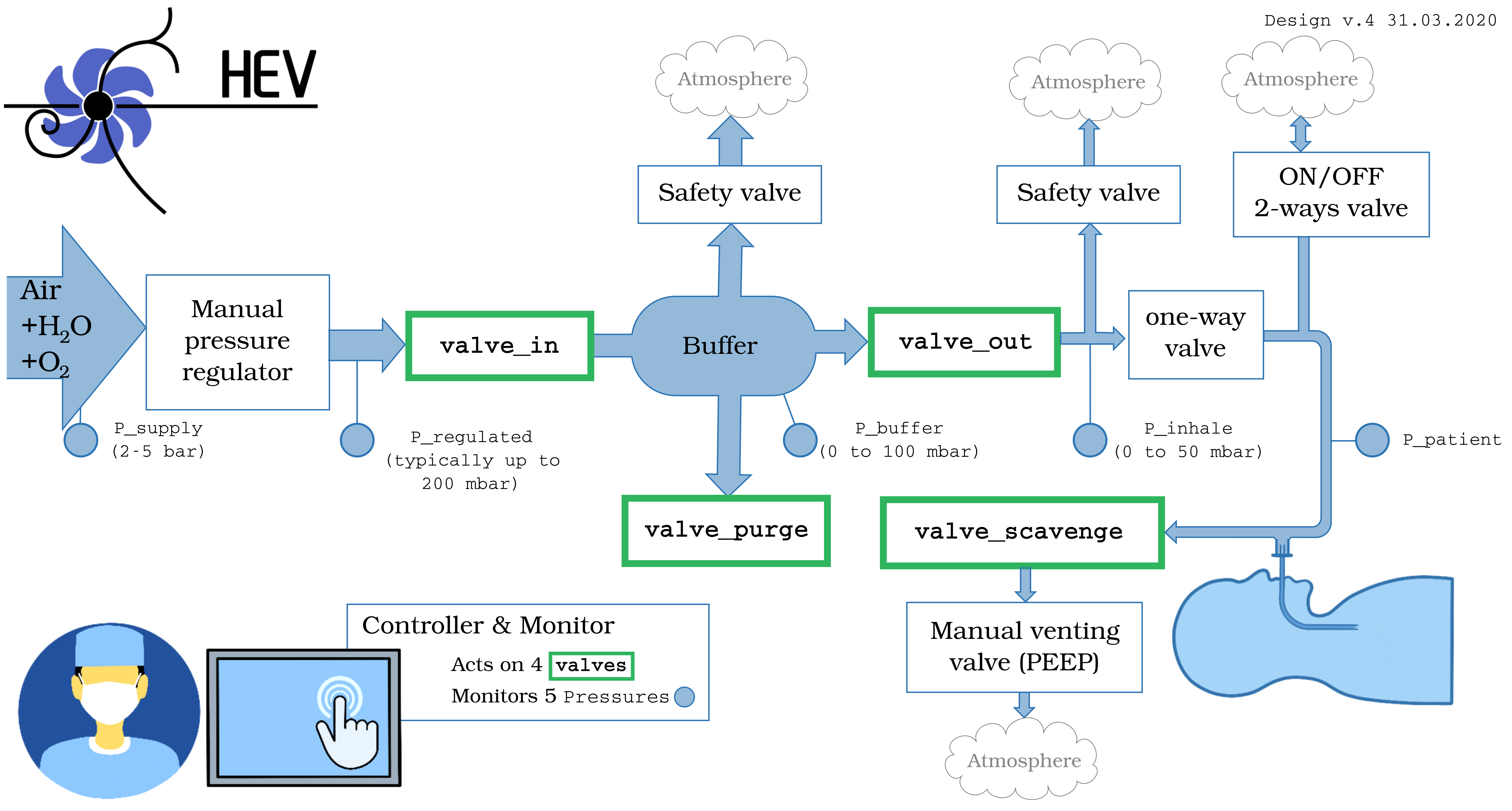}
  
  \caption{Conceptual design of the HEV ventilator.}
  \label{fig:hev_concept}
  \end{center}
\end{figure}

The conceptual schematic is shown in figure~\ref{fig:hev_concept}.  The unit takes as input the standard compressed or mixed air supply available in hospitals, in such a way that one supply could be connected to several units.  We expect that typically the pressure supplied will be between 2 and 5 bar. The connections presented by the unit to external input/outputs will follow hospital standards. The supply pressure is reduced by a pressure regulator to approximately 200 mbar.  The system concept is based around a buffer volume of approximately 2 litres. The filling of this buffer is controlled by the input valve ({\tt valve\_in}). By controlling of the opening time, one can achieve the desired target pressure in the buffer after which the valve ({\tt valve\_in}) is shut. This buffer filling occurs during the expiratory part of the breath cycle. If the buffer pressure is within tolerance of the required pressure, the output valve ({\tt valve\_out}) is then opened, initiating the respiratory cycle. The respiratory rate, inspiratory time (corresponding to the open time of {\tt valve\_out}) and pause time are all controllable. If a PEEP pressure is set, then the pressure in the lungs will have the minimum of the PEEP pressure.

We define $P_1$ as the obtained pressure in the buffer before the inhalation cycle starts, and $P_2$ as the pressure in the buffer at the end of the inhalation cycle, when the valve is closed. The volume of air taken by the patient (tidal volume) can then be calculated knowing  $P_1$, $P_2$ and the fixed volume of the buffer and tubes.

During operation all parameters are monitored. If the minute volume drops below the programmed value, the pressure $P_1$ can be increased in units of 1-2~mbar, until the target minute volume is achieved or the maximum pressure allowed is reached.

For the CPAP operation mode all valves are left open and a manual regulator setting is set to a fixed low pressure.

The patient is directly protected from over-pressure via the safety valve, which will open at 80~cm~$\rm H_2 O$.  In addition, the pressure  sensor in the buffer ({\tt P\_buffer}) will continuously check for over-pressure. In case of over-pressure in the buffer, an electro-valve ({\tt valve\_purge}) will be open to purge its contents and refill it to the correct level. In case of failure of output valve of the buffer ({\tt valve\_out}) or a powercut, the ON/OFF two say valve will connect the patient to the atmosphere allowing non-assisted breathing. %An additional valve is installed that opens to air in case the output valve of the buffer mechanically fails in a permanently closed state, allowing autonomous breathing, should the patient be capable of this.
During the expiratory cycle, the valve\_scavenger opens allowing the air from the lungs to flow to the scavenging system.

\begin{figure}[ht]
\begin{center}
  \includegraphics[width=0.95\linewidth]{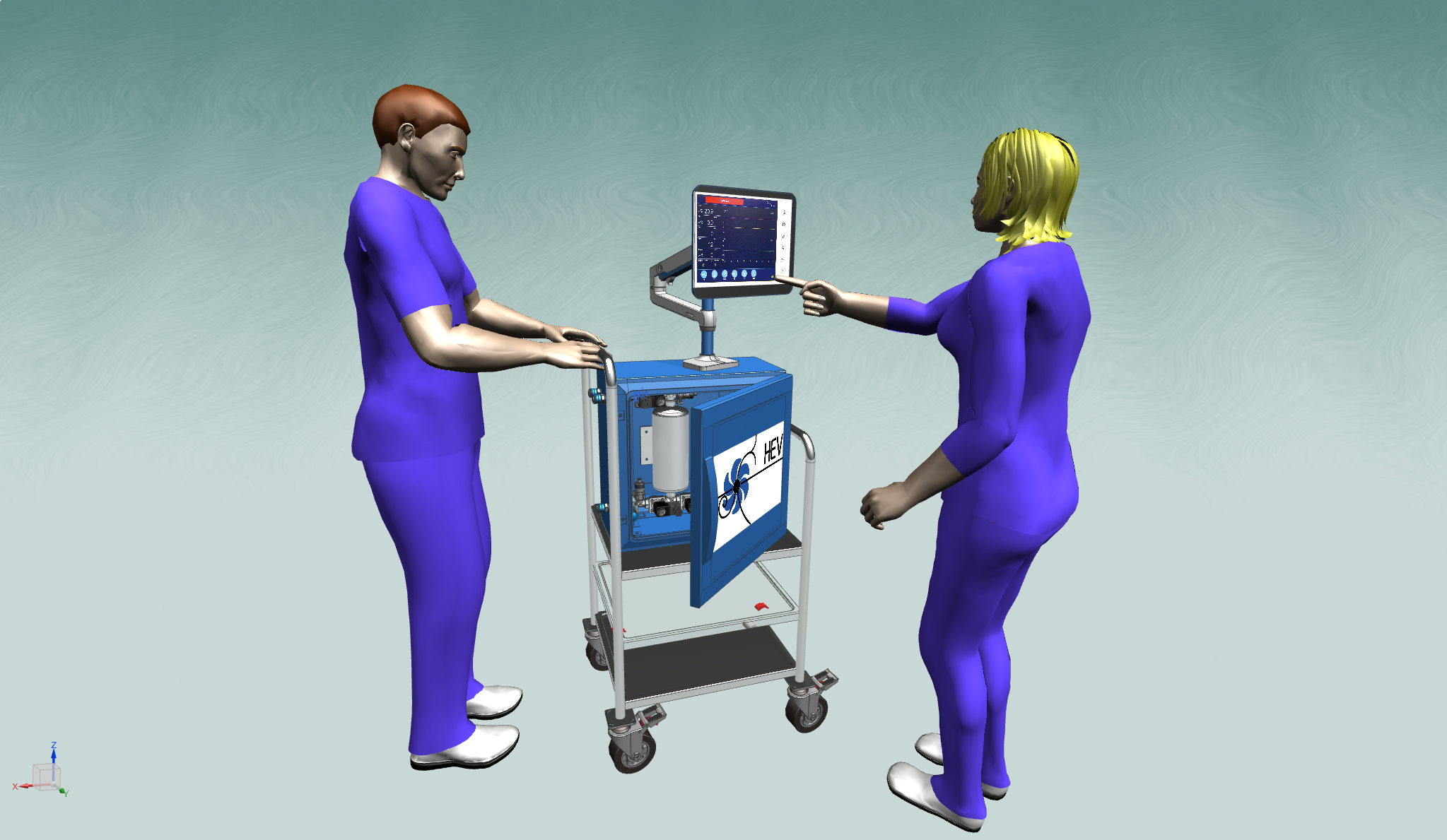}
  \caption{Preliminary CAD model of the HEV proposed design.  The overall dimensions can be seen, together with the placement of the major components in the HEV cabinet.}
  \label{fig:mech_render}
  \end{center}
\end{figure}

The mechanical design will be based on this concept and will result in a unit with approximate size $500 \times 500 \times 350$~mm. The unit dimensions and the positioning of the main components are indicated in figure~\ref{fig:mech_render}.

\section{Specifications}

The HEV ventilator will be designed to the following specifications:
\begin{itemize}
    \item Working Pressure: Up to 50~cm~$\rm H_2 O$.
    \item PRVC peak pressure limit set to 35~cm~$\rm H_2 O$ by default with an option to increase this in exceptional circumstances and by positive decision and action by the user.
    \item Operation modes: PRVC, SIMV-PC, CPAP, as defined above.
    \item Exhaust mode: PEEP available with a set range between 0 and 5~cm~$\rm H_2 O$.
    \item Minute volume flow capability: Up to 20 litres/min. 
    \item Inspiratory  flow capability: Up to 120 litres/min. 
    \item Respiratory rate: 10--30 breaths/min.
    \item Inspiratory:Expiratory ratio; 1:2 will be provided as standard, and the unit will be adjustable in the range 1:1--1:3. 
    \item Tidal volume setting to be provided in the range 250--800~ml in steps of 50~ml.
    \item Gas and Power Supply Inlet: Set according to the MHRA standards~\cite{mhra}.
\end{itemize}

\section{Control and User Interface}

\begin{figure}[ht]
\begin{center}
  \includegraphics[width=0.85\linewidth]{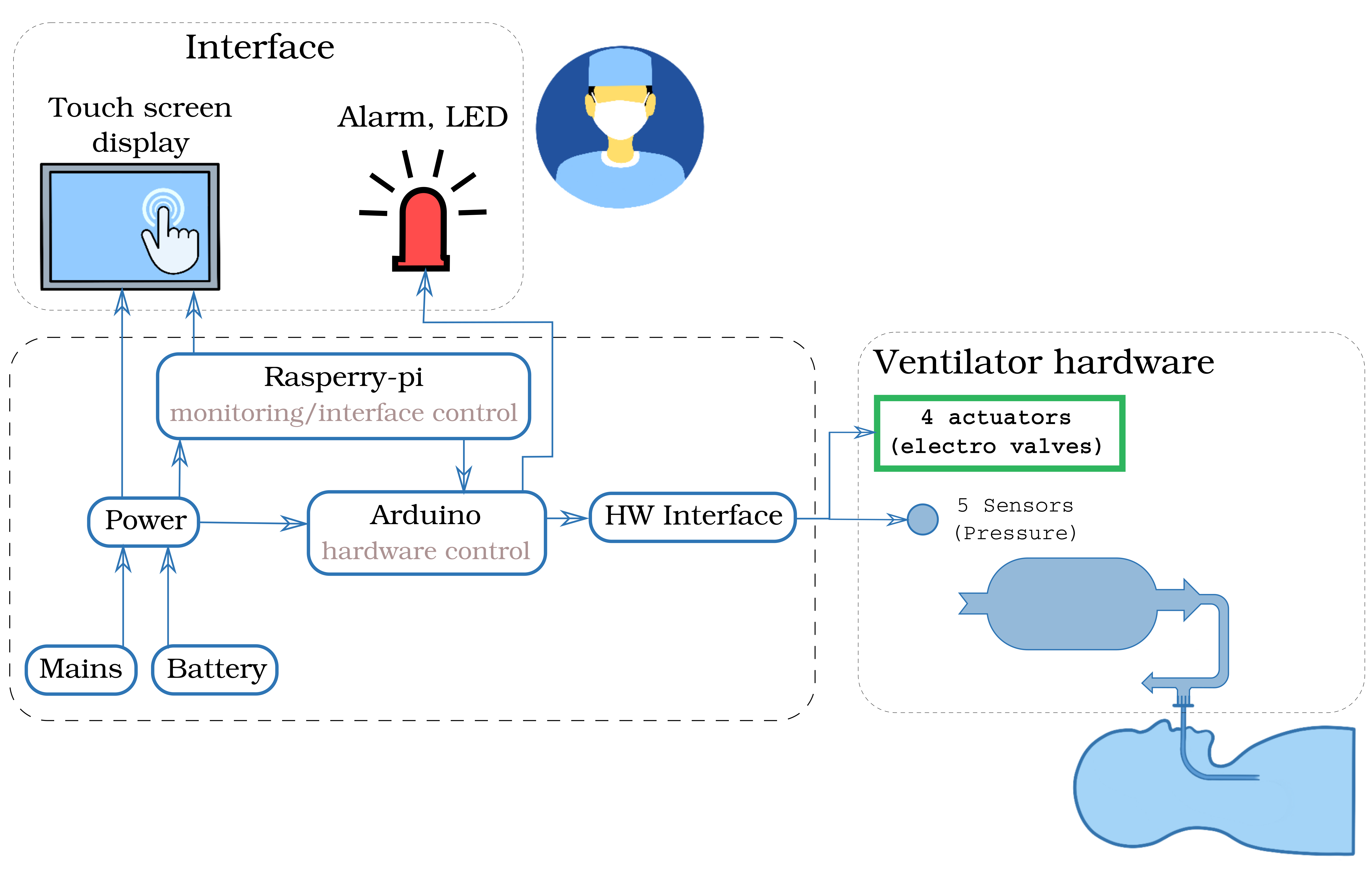}
  \caption{Conceptual layout of the controls and interface}
  \label{fig:control_diagram}
  \end{center}
\end{figure}

The control concept, based around the embedded controller receiving the signals from the sensors and valves, and the touchscreen interface to the clinician, is illustrated in figure~\ref{fig:control_diagram}. 
The user interface will consist of an intuitive panel which displays the needed information and controls in a dynamic and continuous fashion.  The controls and alarms described here are extremely preliminary and simply give an idea of the system functionality.  The finalised list will emerge during the prototyping stages.

The following items will be included in the display:
\begin{itemize}
    \item Ventilation mode.
    \item Working pressure, corresponding to the manually set and monitored input pressure to the unit.
    \item Inspiratory Minute Volume setting.
    \item Breaths per minute setting.
    \item Tidal volume display, based on previous two parameters.
    \item Inspiration time setting.
    \item Pause time setting.
    \item Expiration time display based on previous two parameters.
    \item Expired minute volume.
    \item Airway pressure display based on the reading of $P_2$.
    \item PEEP setting.
    \item Trigger sensitivity to patient-initiated breath.
\end{itemize}

The following items will be plotted as a continuous graphical display:
\begin{itemize}
    \item Volume vs time.
    \item Pressure vs time.
    \item Flow vs time.
\end{itemize}

%The following items will be displayed as alarms:

%\begin{itemize}
%    \item Push Button manual alarm
%    \item Gas supply alarm
%    \item Apnea alarm
%    \item Expired minute volume
%    \item Upper pressure limit
%    \item Lower failure
%\end{itemize}

The HEV will pay particular attention to the MV leak alarm which is a standard hospital setting alarm, but of crucial importance in the case of a highly infectious disease such as COVID-19 to protect staff and patients.  It will trigger with high sensitivity on any leak in a circuit, which could come from the balloon in the patient airway leaking, a disconnection or a poorly tightened tube and so on. In the current version the $\rm O_2$ concentration is not measured, but this is being investigated actively. The display format will be similar to the typical display seen in a hospital setting.

In pressure support modes the ventilator will go into an alarm state and fail-safe to mandatory ventilation in case the patient-initiated breaths cease.  For all modes alarms are given if breath rate or minute volumes fall below acceptable levels.

The following items will be displayed as audible and visible alarms:
\begin{itemize}
    \item Push Button manual alarm.
    \item Gas supply alarm.
    \item Apnea alarm.
    \item Expired minute volume.
    \item Upper pressure limit.
    \item Power failure.
\end{itemize}

%In the current version the $\rm O_2$ concentration is not measured, but this is being investigated actively.  The format will be similar to the typical display seen in a hospital setting.

\subsection{Technology choice}

Several solutions are currently being considered for the embedded controller including Arduino, Raspberry Pi, and ESP32.  Each have their strengths and weaknesses. Raspberry Pi performs better for a human interface and greater graphical power, and support for HDMI which allows for larger displays.  

Raspberry Pi lacks ADCs which precludes the use of some devices.  On the other hand, the requirement of an operating system adds a complexity that may reduce overall stability.

Arduino and ESP32 have ADC capability and Wifi in several variants, but mostly the graphical add-ons tend to be of a smaller form-factor (5 inches or less) which may not be ideal for hospital use.  The ESP32 is an Arduino compatible device with additional CPU power and memory. The parts supply is being investigated and the availability may be a significant deciding factor in the technology choice. Additionally, given the potential use in several geographical locations, different variants may be developed to adapt to locale.

The current baseline design considers Arduino for controls and Raspberry Pi for monitoring.  The combination of the two offers the most power, and many developers already have one or both of these devices in hand whilst working from home.  Code will be written in a portable manner should a change of platform be required.

\begin{figure}[hbt]
\begin{center}
  \includegraphics[width=\textwidth]{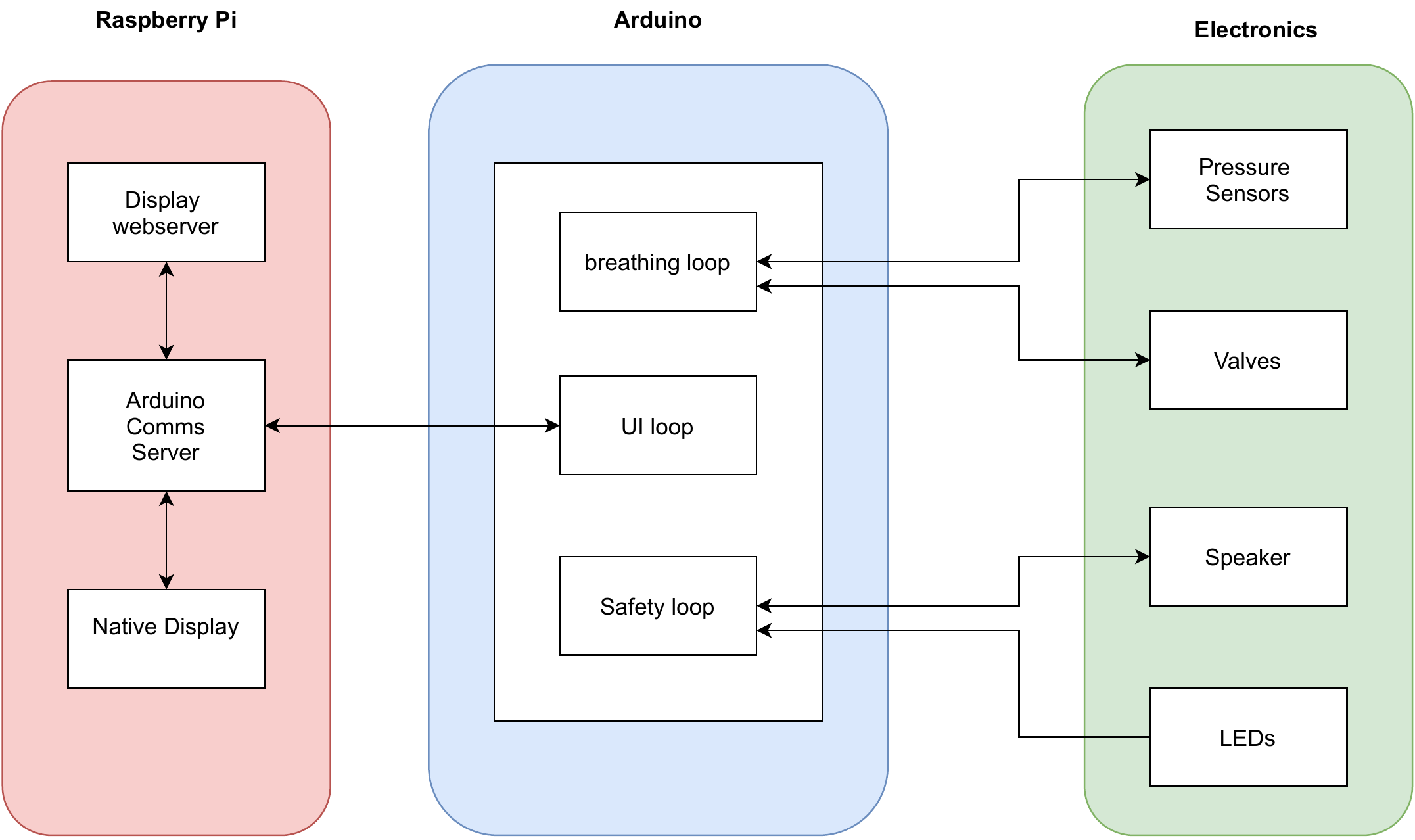}
  \caption{Core software components and their interconnections.  Raspberry Pi is used for
  user interface, Arduino for primary control and connections to the electronics.  }
  \label{fig:hev-sw}
  \end{center}
\end{figure}

\section{Typical parts required}
For the mechanical structure the aim is to use as many standardised parts as possible. A list of the typical parts required includes: 
\begin{itemize}
    \item Solenoid valve with ID bigger then 22~mm. This is the assumed diameter of the hose to the patient. 
    \item Fast acting solenoid valve for inlet of air into buffer volume. 
    \item Solenoid valve for purging. It could be low volume, that in case of over-pressure only some air is released. 
    \item 2 litre buffer container. 
    \item Pressure regulator from 2-5~bar to up to 200~mbar. 
    \item 5 pressure sensors\footnote{The pressure sensor Panasonic PS-A (ADP5) is currently being tested.}.
\end{itemize}

For the embedded controller similar concepts apply.  A typical part list could include:
\begin{itemize}
    \item 1 Arduino and 1 Raspberry Pi per unit.
    \item 1 touchscreen, preferably 15", also 10" could be acceptable.  If the touchscreens are not available an option could be a standard PC monitor with an interface appropriate for the hospital setting i.e. with buttons.  Medical touchscreens could be an option but are likely to be more expensive.
    \item Typical small supplies such as 5V power supplies, power supplies for the valves, USB cables (1 standard, 2 micromax).
    \item Optional ethernet cable.
\end{itemize}

\section{Prototyping}

\begin{figure}[hbt]
\begin{center}
  \includegraphics[width=0.3\linewidth]{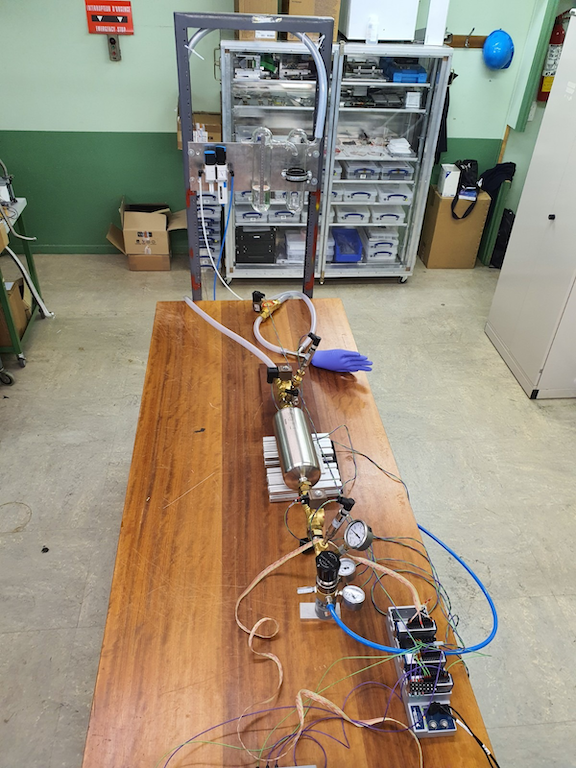}
  \includegraphics[width=0.3\linewidth]{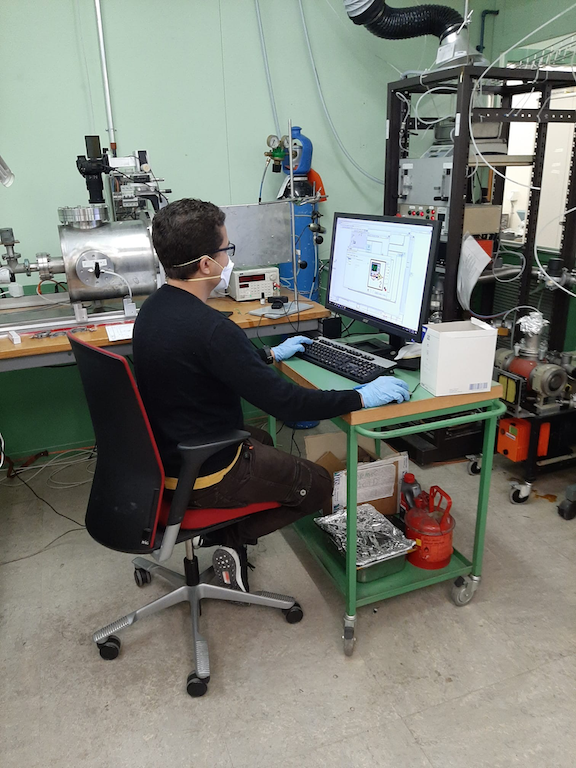}
  \caption{Prototyping the HEV ``buffer concept".  The demonstrator system is developed with the buffer concept to demonstrate the ``breathing" and flow capabilities.  This demonstrator is built with in-house parts and looks mechanically very different to the final system.  Control is provided via LabView, whereas in the final system it will be via an embedded controller.}
  \label{fig:first_prototype_buffer}
  \end{center}
\end{figure}

The first stage of prototyping is to demonstrate that the working principle of an alternatively filled and emptied buffer is sound, and allows the ventilator to operate within the required ranges of pressure and time. This was achieved on 27$^{\rm th}$ March 2020 with a first demonstrator. A photograph of the bench test stand is shown in figure~\ref{fig:first_prototype_buffer}.  Note that the demonstrator was built with readily available parts, and therefore the mechanical look of the device seen in these photographs is completely different to the final design.  Based on the experience with operating the demonstrator, the desired physical characteristics of the pressure regulators, valves and pressure sensors could be refined, and the system is now in prototyping stage.

\section{Route to deployment}

The deployment of the device will pass through the following stages:
\begin{enumerate}
    \item In the first phase the concept will be demonstrated with a working unit prepared using the available hardware and with a control system implemented with standard software.
    \item Consultations on testing and validation are taking place with hospital clinicians, both in the CERN region and beyond.
    \item Based on the experience of the demonstrator unit and the consultation with the clinicians, a prototype will be built and tested possibly with standard test facilities for ventilators, available in hospitals. 
    \item Control software will be developed, able to run in embedded microprocessors and with a dedicated graphical user interface allowing clinicians access to the settings and providing alarms in a mode familiar to the operators.  This will shadow the standard software development in such a way that the hospital deployment gives identical functionality.
    \item As far as production is concerned, it is foreseen, on the one hand, to enable this through providing partner academic institutions with the detailed design (as soon as available following completion of testing and validation), for these institutions to follow up in accordance with local possibilities and standards; on the other hand, directly through industry, for which purpose contacts have been established with the WHO, on the basis of the Cooperation Agreement in place between CERN and WHO.
\end{enumerate}

%Phase 1 demonstrate prototype, phase 2 deploy at hospital in consultation with clinicians, phase 3 distribute blueprint to network of academic institutes with local hospital contacts

%https://www.overleaf.com/project/5e7db3c95fec380001c690b1

\section*{Acknowledgements}

\noindent We express our gratitude to
Paolo Chiggiato, Beniamino Di Girolamo, Walid Fadel, Doris Forkel-Wirth, Andre Henriques, Christian Joram, Rolf Lindner, David Reiner, Burkhard Schmidt, Francois Vasey from CERN, Geneva, Switzerland,
% comment from here to ...
Simon Cohen from Monash Children's Hospital, Melbourne, Australia, Gordon Flynn and David Reiner from The Canberra Hospital, Canberra, Australia, Hamish Woonton from Dandenong Hospital and Monash Health, Melbourne, Australia, Bruce Dowd from Prince of Wales Hospital, NWS, Australia, Lise Piquilloud Imboden and Patrick Schoettker from Centre Hospitalier
Universitaire Vaudois, Lausanne, Switzerland, Georg M\"annel and Philipp Rostalski from Universit\"at zu L\"ubeck, L\"ubeck, Germany,
Dr Roosens, University Hospital, Ghent,
% ... there to remove the doctors from acknoledgement (if passing them as authors)
for many illuminating discussions and much practical support.

\addcontentsline{toc}{section}{References}
%\setboolean{inbibliography}{true}
%\bibliography{main}%,standard,LHCb-PAPER,LHCb-DP,LHCb-TDR}
\printbibliography

\newpage
% $Id: LHCb_authorlist.tex 78711 2015-08-06 07:54:32Z apuignav $
% ===============================================================================
% Purpose: example of authorlist for LHCb template
% Author:
% Created on: 2009-09-24
% ===============================================================================

%\documentclass[a4paper]{article}
%\setlength{\oddsidemargin}{0cm}
%\setlength{\evensidemargin}{0cm}
%\setlength{\textwidth}{16.5cm}
%\setlength{\parindent}{0cm}
%\begin{document}
\centerline{\large\normalfont\bfseries HEV collaboration}
\begin{flushleft}
\small
%-- LHCb Authorlist, Example typesetting
%-- 
J.~Buytaert$^{1,*}$, 
A.~Abed~Abud$^{1,2}$, 
K.~Akiba$^{3}$, 
A.~Bay$^{10}$, 
C.~Bertella$^{1}$,
T.~Bowcock$^{2}$, 
W.~Byczynski$^{1,8}$,
V.~Coco$^{1}$, 
%S.~Cohen$^{12}$,
P.~Collins$^{1,*}$, 
O.~Augusto~De~Aguiar Francisco$^{1}$, 
%%%B.~Di~Girolamo$^{1}$, 
N.~Dikic$^{1}$, 
R.~Dumps$^{1}$, 
P.~Durante$^{1}$,  
A.~Fern\'andez Prieto$^{11}$,
%G.~Flynn$^{13}$, 
V.~Franco Lima$^{2}$,
R.~Guida$^{1}$, 
K.~Hennessy$^{2}$,
D.~Hutchcroft$^{2}$, 
S.~Ilic$^{7}$, 
A.~Jevtic$^{7}$, 
%%%C.~Joram$^{1}$, 
K.~Kapusniak$^{1}$, 
E.~Lemos~Cid$^{11}$,
J.~Lindner$^{9}$, 
M.~Milovanovic$^{6}$,
D.~Murray$^{4}$, 
I.~Nasteva$^{5}$, 
N.~Neufeld$^{1}$, 
%L.~Piquilloud Imboden$^{17}$
X.~Pons$^{1}$, 
%D.~Reiner$^{13}$, 
F.~Sanders$^{3}$, 
%%%B.~Schmidt$^{1}$, 
%P.~Schoettker$^{17}$
R.~Schwemmer$^{1}$,
P.~Svihra$^{4}$. %, instead if not the last
%H.~Woonton$^{14,15}$.
\bigskip\newline{\it
%B.~Dowd$^{16}$,
\footnotesize

% note LHCb institute address copied from a recent paper
% https://journals.aps.org/prl/pdf/10.1103/PhysRevLett.124.082002

$ ^{1}$European Organization for Nuclear Research (CERN), Geneva, Switzerland\\
$ ^{2}$Oliver Lodge Laboratory, University of Liverpool, Liverpool, United Kingdom\\
$ ^{3}$Nikhef National Institute for Subatomic Physics, Amsterdam, Netherlands\\
$ ^{4}$Department of Physics and Astronomy, University of Manchester, Manchester, United Kingdom\\
$ ^{5}$Universidade Federal do Rio de Janeiro (UFRJ), Rio de Janeiro, Brazil\\
$ ^{6}$Deutsches Elektronen-Synchrotron (DESY), Platanenallee 6, 15738 Zeuthen, Germany \\
$ ^{7}$University of Nis, Univerzitetski trg 2, Niš 18000, Serbia \\
$ ^{8}$Tadeusz Kosciuszko Cracow University of Technology, Cracow, Poland \\
$ ^{9}$University of Applied Sciences Offenburg, Offenburg, Baden-Wuerttemberg,  Germany \\
$ ^{10}$Institute of Physics, Ecole Polytechnique F\'ed\'erale de Lausanne (EPFL), Lausanne, Switzerland \\
$ ^{11}$Instituto Galego de Física de Altas Enerxías (IGFAE), Universidade de Santiago de Compostela, Santiago de Compostela, Spain \\
}
%$ ^{12}$ Monash Children's Hospital, Melbourne, Australia \\
%$ ^{13}$ The Canberra Hospital, Canberra, Australia\\
%$ ^{14}$Monash Health, Melbourne, Australia\\
%$ ^{15}$Dandenong Hospital, Melbourne, Australia\\
%$ ^{16}$ Prince of Wales Hospital, New South Wales, Australia\\
%$ ^{17}$Centre Hospitalier Universitaire Vaudois, Lausanne, Switzerland Hospital \\
$ ^{*}$Corresponding authors: Jan.Buytaert@cern.ch and Paula.Collins@cern.ch
%-- 
%-- 
\end{flushleft}
%\end{document}

\end{document}